\documentclass[a4paper,11pt]{article}
\usepackage[hmargin=3cm,vmargin=3cm]{geometry}
\usepackage{graphicx, amssymb, textcomp, hyperref, amsmath}
\numberwithin{equation}{section}
\title{\bf Quasi-exactly solvable relativistic soft-core Coulomb models}
\author{\Large Davids Agboola \footnote{d.agboola@maths.uq.edu.au}~ and~ Yao-Zhong Zhang \footnote{yzz@maths.uq.edu.au}}
\date{\it School of Mathematics and Physics, The University of Queensland, \\Brisbane, QLD 4072, Australia}
\begin{document}
\maketitle
\vspace{0.5in}
\noindent {\bf Abstract}~By considering a unified treatment, we present quasi exact polynomial solutions to both the Klein-Gordon and Dirac equations with the family of soft-core Coulomb potentials $V_q(r)=-Z/\left(r^q+\beta^q\right)^{1/q}$, $Z>0$, $\beta>0$, $q\geq 1$. We consider cases $q=1$ and $q=2$ and show that both cases are reducible to the same basic ordinary differential equation. A systematic and closed form solution to the basic equation is obtain using the  Bethe ansatz method. For each case, the expressions for the energies and the allowed parameters are obtained analytically and the wavefunctions are derive in terms of the roots of a set of Bethe ansatz equations.

\vspace{.5in}
\section{Introduction}
The search for the exact solutions corresponding to the eigenvalue equations of the form $H\Psi=E\Psi$ is one of the major research interests in quantum mechanics. As such solutions play a vital role in the explanation of many physical phenomena, increasing efforts are being made to present both physically relevant and mathematically valid solutions. For standard quantum-mechanical systems, such as the hydrogen atom and the harmonic oscillator, solutions to the eigenvalue equation are usually constructed by transforming the original differential equations into a differential equation whose solution are known in terms of  a certain special function. Thus in most cases, it is not uncommon to have the required solution in form of the hypergeometric function, which in turn has led to a deeper understanding of these systems under both free and spatially confined conditions.

In particular, solutions to relativistic equations play a very important role in many aspects of of modern physics. For instance, the Dirac equation has been used to explain the anti-nucleon bound in a nucleus \cite{1}, deformed nuclei \cite{2}, super deformation \cite{3} and has been used to establish an effective nuclear shell model scheme \cite{4}-\cite{6}; while the Klein-Gordon equation has been used in describing a wide variety of phenomena, which include classical wave systems, such as the displacement of a string attached to an elastic bed \cite{7}, and quantum systems based on scalar field theories \cite{8}. 

On the  other hand, the potentials 
\begin{equation}\label{eq:1}
V_q(r)=-\frac{Z}{\left(r^q+\beta^q\right)^{1/q}}.
\end{equation}
which represent a family for the soft-core Coulomb potentials have been useful  in atomic and molecular physics. The nonrelativistic bound states have been obtained in terms of the parameters: the coupling $Z>0$, the cut-off parameter $\beta>0$, the power parameter $q\geq 1$. The potential $V_1$ represents the potential due to a smeared charge and is useful in describing mesonic atoms. The potential $V_2$ is similar to the shape of the potential due to a finite nucleus and experienced by muon in a muonic atom. Also, several applications of the $V_2$ potential has been made through model calculations corresponding to the interation of intense laser fields with atom \cite{14}-\cite{17}. The parameter $\beta $ can be related to the strenght of the laser field, within the range of $\beta=20-40$ covering the experimental laser field strength \cite{14}. 

Due to the limited applications of exactly solvable systems in quantum mechanics, recent attentions have been on  the systems with partially solvable spectral. Such systems are said to be Quasi-Exactly Solvable (QES). Thus a quantum mechanical system is called quasi-exactly solvable, if only a part of the eigenvalues and corresponding eigenvectors can be obtained exactly \cite{18}-\cite{21}.  
Solutions to QES system can be constructed in a number ways. For instance, the supersymmetric speration method which has been used to obtain the solutions to several two-dimensional partially solvable models \cite{iofe1,iofe2}. Also, one can separate the asymptotic behaviours of the system to get an equation which can be expanded as a power series of the basic variable. This equation unlike an exactly solvable equation with two-step recursive relations, possess a three-step recursive relation for the coefficients of the power series. The complexity of the recursive relations does not allow one to guarantee the square integrability property of the wavefunction. However, by choosing a polynomial wavefunction, one can terminate the series at a certain order and then impose a sufficient condition for the normalization. By so doing, exact solutions to the system can be obtained, but only for certain energies and for special values of the parameters of the problem.

Solutions to QES systems have mostly been discussed in terms of the the recursion  relations of the power series coefficients, which is mostly expressed in terms of the (generalized) Heun differential equations \cite{10,11}, except for few cases where the factorization method has been used \cite{22}. Although the solutions obtained in connection with the Heun equations are exact but the procedures involved are quite ambiguous, thus expunging the closed form of the solutions. In a series of recent studies \cite{23}-\cite{DA12}, the Bethe Ansatz Method (BAM) has been used in obtaining the solutions to QES systems. This method did not only yield exact solutions, it also preserve the closed form representation of the solutions. For instance, the BAM has been used to obtain the solutions of QES difference equation \cite{23} and the exact polynomial solutions of general quantum non-linear optical models \cite{24,25}. Very recently, the method has also been used to obtain the exact solutions for a family of spin-boson systems \cite{26}.

The aim of this paper is to extent the works of \cite{9}-\cite{13}, by presenting relativistic treatments of potential \eqref{eq:1} using the  BAM. The work is arranged as follows: in section 2, we present the Dirac and Klein-Gordon equations with the soft-core potentials and reduce each equation to a QES equation using appropriate transformations. In section 3, we present the solutions to the basic underlying differential equation for both models using the BAM. The application of the main results to obtain the energies and wavefunction for each model is presented in section 4, while in section 5, we discuss the hidden Lie algebraic structure of the models. Final concluding remarks are given in section 6.

\section{Relativistic Soft-core Coulomb Models}
In this section we give a description of both the Dirac and Klein-Gordon cases. The corresponding equations are reduced to a QES equation using appropriate transformation. Except otherwise stated, we employ the natural units $\hbar=c=e=1$ throughout the paper.
\subsection{The Dirac equation with the soft-core Coulomb potentials}
The Dirac equation for a single-nucleon with mass $\mu$ moving in a spherically symmetric attractive scalar $S_q(r)$ and repulsive vector $V_q(r)$ soft-core Coulomb potentials is written as \cite{28}-\cite{30}  
\begin{equation}\label{eq:2}
H\Psi_{q,n}{({\bf r})}=E_{q,n}\Psi_{q,n}({\bf r})\hspace{.2in}\mbox{where}\hspace{.1in}H=\sum_{j=1}^3\hat{\alpha}_jp_j+\hat{\beta}[\mu+S_q(r)]+V_q(r),
\end{equation}
$E_{q,n}$ is the relativistic energy, 
\begin{equation}\label{eq:2.1}
V_q(r)=-\frac{Z_v}{\left(r^q+\beta^q\right)^{1/q}}~~~~~~\mbox{and}~~~~~~ S_q(r)=-\frac{Z_s}{\left(r^q+\beta^q\right)^{1/q}},
\end{equation}
$\{\hat{\alpha}_j\}$ and $\hat{\beta}$ are Dirac matrices defined as 
\begin{equation}\label{eq:3}
\hat{\alpha}_j=\begin{pmatrix}
0&\hat{\sigma}_j \\
\hat{\sigma}_j&0

\end{pmatrix}\hspace{.3in}\hat{\beta}=\begin{pmatrix}
\bf{1}&0 \\
0&\bf{-1}

\end{pmatrix} 
\end{equation}
where $\hat{\sigma}_j$ is the Pauli's $2\times 2$ matrices and $\hat{\beta}$ is a $2\times 2$ unit matrix, which satisfy anti-commutation relations
\begin{equation}\label{eq:4}
\begin{array}{lrl}
\hat{\alpha}_j\hat{\alpha}_k+\hat{\alpha}_k\hat{\alpha}_j&=&2\delta_{jk}\bf{1}\\
\hat{\alpha}_j\hat{\beta}+\hat{\beta}\hat{\alpha}_j&=&0\\
{\hat{\alpha}_j}^2=\hat{\beta}^2&=&\bf{1}
\end{array}
\end{equation}
and $p_j$ is the three momentum which can be written as
$$p_j=-i\partial_j=-i\frac{\partial}{\partial x_j} \hspace{.2in} 1\leqslant j\leqslant 3.$$ 
The orbital angular momentum operators $L_{jk}$, the spinor operators $S_{jk}$ and the total angular momentum operators $J_{jk}$ can be defined as follows:
$$L_{jk}=-L_{jk}=ix_j\frac{\partial}{\partial x_k}-ix_k\frac{\partial}{\partial x_j},\hspace{.2in} S_{jk}=-S_{kj}=i\hat{\alpha}_j\hat{\alpha}_k/2,\hspace{.2in} J_{jk}=L_{jk}+S_{jk}.$$
\begin{equation}\label{eq:5}
L^2=\sum_{j<k}^3L^2_{jk},\hspace{.2in}S^2=\sum_{j<k}^3S^2_{jk},\hspace{.2in}J^2=\sum_{j<k}^3J^2_{jk}, \hspace{.2in} 1\leqslant j< k\leqslant 3.
\end{equation}
For a spherically symmetric potential, total angular momentum operator $J_{jk}$ and the spin-orbit operator $\hat{K}=-\hat{\beta}(J^2-L^2-S^2+1/2)$ commutate with the Dirac Hamiltonian. For a given total angular momentum $j$, the eigenvalues of $\hat{K}$ are $\kappa=\pm(j+1/2)$; $\kappa=-(j+1/2)$ for aligned spin $j=\ell+\frac{1}{2}$ and $\kappa=(j+1/2)$ for unaligned spin $j=\ell-\frac{1}{2}$. Moreover, the spin-orbital quantum number $\kappa$ is related to the orbital angular quantum number $\ell$ and the pseudo-orbital angular quantum number $\tilde{\ell}=\ell+1$ by the expressions $\kappa(\kappa+1)=\ell(\ell+1)$ and $\kappa(\kappa-1)=\tilde{\ell}(\tilde{\ell}+1)$ respectively for $\kappa=\pm 1, \pm 2,\dots$. The spinor wave functions can be classified according to the  radial quantum number $n$ and the spin-orbital quantum number $\kappa$ and can be written using the Dirac-Pauli representation
\begin{equation}\label{eq:6}
\Psi_{q,n}({\bf r})=\frac{1}{r}\left(\begin{array}{lll}
F_{q,n}(r)Y_{jm}^\ell\left(\theta,\phi\right)\\\\
iG_{q,n}(r)Y^{\tilde{\ell}}_{jm}\left(\theta,\phi\right)
\end{array}\right)
\end{equation}
where $F_{q,n}(r)$ and $G_{q,n}(r)$ are the radial wave function of the upper- and the lower-spinor components respectively, $Y_{jm}^\ell\left(\theta,\phi\right)$ and $Y^{\tilde{\ell}}_{jm}\left(\theta,\phi\right)$ are the spherical harmonic functions coupled with the total angular momentum $j$. The orbital and the pseudo-orbital angular momentum quantum numbers for spin symmetry  $\ell$ and  and pseudospin symmetry $\tilde{\ell}$ refer to the upper- and lower-component respectively.  

Substituting Eq.\,\eqref{eq:6} into Eq.\,\eqref{eq:2}, and separating the variables we obtain the following coupled radial Dirac equation for the spinor components:
\begin{equation}\label{eq:7a}
\left(\frac{d}{dr}+\frac{\kappa}{r}\right)F_{q,n}(r)=[\mu+E_{q,n}-\Delta(r)]G_{q,n}(r)
\end{equation}
\begin{equation}\label{eq:7b}
\left(\frac{d}{dr}-\frac{\kappa}{r}\right)G_{q,n}(r)=[\mu-E_{q,n}+\Sigma(r)]F_{q,n}(r)
\end{equation}
where $\Sigma_q(r)=V_q(r)+S_q(r)$ and $\Delta_q(r)=V_q(r)-S_q(r)$. Using Eq.\,\eqref{eq:7a} as the upper component and substituting into Eq.\,\eqref{eq:7b}, we obtain the following second order differential equations
\begin{equation}\label{eq:8a}
\small\left[\frac{d^2}{dr^2}-\frac{\kappa(\kappa+1)}{r^2}-[\mu+E_{q,n}-\Delta_q(r)][\mu-E_{q,n}+\Sigma_q(r)]+\frac{\frac{d\Delta_q(r)}{dr}\left(\frac{d}{dr}+\frac{\kappa}{r}\right)}{[\mu+E_{q,n}-\Delta_q(r)]}\right]F_{q,n}(r)=0, 
\end{equation}
\begin{equation}\label{eq:8b}
\small\left[\frac{d^2}{dr^2}-\frac{\kappa(\kappa-1)}{r^2}-[\mu+E_{q,n}-\Delta_q(r)][\mu-E_{q,n}+\Sigma_q(r)]-\frac{\frac{d\Sigma_q(r)}{dr}\left(\frac{d}{dr}-\frac{\kappa}{r}\right)}{[\mu-E_{q,n}+\Sigma_q(r)]}\right]G_{q,n}(r)=0. 
\end{equation}
To solve these equations, we employ the pseudospin symmetry limit, in which the Dirac Hamiltonian is invariant under the $SU(2)$ algebra with the attractive scalar potential and the repulsive vector potential having nearly equal magnitudes, i.e. $V_q(r)\sim S_q(r)$ or $V_q(r)+S_q(r)=C_q$, $C_q$ being the pseudospin constants. This implies $\frac{d\Sigma_q(r)}{dr}=0$ and hence Eq.\,\eqref{eq:8b} takes a simple form 
\begin{equation}\label{eq:9}
\left\{\frac{d^2}{dr^2}-\frac{\kappa(\kappa-1)}{r^2}-\left[\mu+E_{q,n}-\Delta_q(r)\right]\left[\mu-E_{q,n}+C_q\right]\right\}G_{q,n}(r)=0. 
\end{equation}
Details of the concepts and applications of pseudospin symmetry can be found in \cite{1, 4, 5, 29} and the references therein. If we take 
\begin{equation}\label{eq:10}
\Delta_q(r)=-\frac{Z_\delta}{\left(r^q+\beta^q\right)^{1/q}}
\end{equation}
where $Z_\delta=Z_v-Z_s$, $Z_v$ and $Z_s$ corresponding to the coupling parameters for the vector and scalar potentials respectively, then Eq.\,\eqref{eq:9} becomes
\begin{equation}\label{eq:11}
G''_{q,n}(r)-\frac{\kappa(\kappa-1)}{r^2}G_{q,n}(r)-\left[\sigma^2+\frac{\gamma Z_\delta}{\left(r^q+\beta^q\right)^{1/q}}\right]G_{q,n}(r)=0
\end{equation}
where 
\begin{equation}\label{eq:12}
\gamma=\mu-E_{q,n}+C_q, \hspace{0.1in}\mbox{and}\hspace{0.1in}\sigma^2=(\mu+E_{q,n})\gamma.
\end{equation}
We introduce the variable $t=\left(r^q+\beta^q\right)^{1/q}$ and then transform Eq.\,\eqref{eq:11} using
\begin{equation}\label{eq:13}
G_{q,n}(t)=\left(t^q-\beta^q\right)^{\kappa/q}e^{-\omega t}\varphi(t)
\end{equation} to have
\begin{equation}\label{eq:14}
\varphi ''(t)+\left[\frac{(q-1)\beta^q+2\kappa t^{q}}{t\left(t^q-\beta^q\right)}-2\omega\right]\varphi '(t)-\left[\frac{\omega(q-1)\beta^q+2\omega\kappa t^{q}}{t\left(t^q-\beta^q\right)}-\omega^2\right.$$
$$\left.+\frac{\gamma Z_\delta}{t^{3-2q}\left(t^q-\beta^q\right)^{2-2/q}}+\frac{\sigma^2}{t^{2-2q}\left(t^q-\beta^q\right)^{2-2/q}}\right]\varphi(t)=0.
\end{equation}

\subsubsection{Case $q=1$} With $q=1$, Eq.\,\eqref{eq:14} becomes
\begin{equation}\label{eq:15}
\varphi ''(t)+2\left[\frac{\kappa}{t-\beta}-\omega\right]\varphi '(t)-\left[\frac{2\omega\kappa}{t-\beta}-\omega^2+\sigma^2+\frac{\gamma Z_\delta}{t}\right]\varphi(t)=0,
\end{equation}
If we choose $\omega=\sigma$, then we have 
\begin{equation}\label{eq:16}
t(t-\beta)\varphi ''(t)+2[(\kappa+\sigma\beta)t-\sigma t^2]\varphi '(t)-[t(2\sigma\kappa +\gamma Z_\delta)-\gamma Z_\delta\beta]\varphi(t)=0
\end{equation}
\subsubsection{Case $q=2$} Similarly, with $q=2$ and $\omega=\sigma$, Eq.\,\eqref{eq:14} becomes
\begin{equation}\label{eq:17}
t(t^2-\beta^2)\varphi ''(t)+\left[-2\sigma t^3+2\kappa t^2+2\sigma\beta^2t+\beta^2\right]\varphi '(t)-\left[(2\sigma\kappa+\gamma Z_\delta)t^2+\sigma^2\beta^2t+\sigma\beta^2\right]\varphi(t)=0
\end{equation}
Eqs.\,\eqref{eq:16} and \eqref{eq:17} are now in form suitable for the application of the BAM.
\subsection{The Klein-Gordon equation with the soft-core Coulomb potentials}
The Klein-Gordon equation for a particle of mass $\mu$ with radially symmetric Lorentz vector and Lorentz scalar soft-core Coulomb potentials, $V_q(r)$ and $S_q(r)$, is given by 
\begin{equation}\label{eq:18}
\left\{-\nabla^2+[\mu+S_q(r)]^2\right\}\Phi_{q,n}({\bf r})=[E_{q,n}-V_q(r)]^2\Phi_{q,n}({\bf r})
\end{equation}
where $E_{q,n}$ and $\Phi_{q,n}$ denote the energy and the wavefunction respectively. We can separating the variables using
\begin{equation}\label{eq:19}
\Phi_{q,n}({\bf r})=\frac{\phi_{n,q}(r)}{r}Y_m^\ell(\Omega)
\end{equation}
where $\phi_{q,n}(r)$ is a radial function and $Y_m^\ell(\Omega)$ is a normalized spherical harmonics with eigenvalues $\ell(\ell+1)$, $\ell=0,1,2,\dots$, and $\Omega$ being the angular components. Inserting Eq.\,\eqref{eq:19} into Eq.\,\eqref{eq:18}, we obtain the radial Klein-Gordon equation
\begin{equation}\label{eq:20}
\frac{d^2\phi_{q,n}(r)}{dr^2}-\left[\frac{\ell(\ell+1)}{r^2}+[\mu+S_q(r)]^2-[E_{q,n}-V_q(r)]^2\right]\phi_{q,n}(r)=0.
\end{equation}
Substituting the $V_q(r)$ and $S_q(r)$, we have
\begin{equation}\label{eq:21}
\frac{d^2\phi_{q,n}(r)}{dr^2}-\left[\frac{\ell(\ell+1)}{r^2}-\frac{\lambda_1}{\left(r^q+\beta^q\right)^{1/q}}+\frac{\lambda_2}{\left(r^q+\beta^q\right)^{2/q}}+\xi^2\right]\phi_{q,n}(r)=0
\end{equation}
where 
\begin{equation}\label{eq:22}
\xi^2=\mu^2-E_{q,n}^2,\hspace{0.1in}\lambda_1=2(\mu Z_s+E_{q,n}Z_v)\hspace{0.1in}\mbox{and}\hspace{0.1in}\lambda_2=Z_s^2-Z_v^2
\end{equation}
with $Z_s$ and $Z_v$ denoting the coupling parameter of the scalar and vector potential respectively.
If we introduce the variable $t=\left(r^q+\beta^q\right)^{1/q}$, followed by the transformation
\begin{equation}\label{eq:23}
\phi_{q,n}(t)=\left(t^q-\beta^q\right)^{\frac{\ell+1}{q}}e^{-\rho t}\chi(t),
\end{equation}
Eq.\,\eqref{eq:21} becomes
\begin{equation}\label{eq:24}
\chi ''(t)+\left[\frac{(q-1)\beta^q+2\nu t^{q}}{t\left(t^q-\beta^q\right)}-2\rho\right]\chi '(t)-\left[\frac{\rho(q-1)\beta^q+2\rho\nu t^{q}}{t\left(t^q-\beta^q\right)}-\rho^2+\xi^2\right.$$
$$\left.-\frac{\lambda_1}{t^{3-2q}\left(t^q-\beta^q\right)^{2-2/q}}+\frac{\lambda^2}{t^{4-2q}\left(t^q-\beta^q\right)^{2-2/q}}\right]\chi(t)=0.
\end{equation}
where $\nu=\ell+1$.
\subsubsection{Case 1: $q=1$}
With $q=1$, Eq.\,\eqref{eq:24} becomes
\begin{equation}\label{eq:25}
\chi ''(t)+2\left[\frac{\nu}{t-\beta}-\rho\right]\chi '(t)-\left[\frac{2\rho\nu}{t-\beta}-\frac{\lambda_1}{t}+\frac{\lambda_2}{t^2}-\rho^2+\xi^2\right]\chi(t)=0,
\end{equation}
Choosing $\rho=\xi$, we then have 
\begin{equation}\label{eq:26}
t^2(t-\beta)\chi ''(t)+\left[-2\xi t^3+(2\nu+2\xi\beta)t^2\right]\chi '(t)-\left[(2\xi\nu-\lambda_1)t^2+(\lambda_1\beta+\lambda_2)t-\lambda_2\beta\right]\chi(t)=0
\end{equation}
\subsubsection{Case 1: $q=2$} 
Similarly from Eq.\,\eqref{eq:24}, if $\rho=\xi$, we have the following equation for $q=2$
\begin{equation}\label{eq:27}
t(t^2-\beta^2)\chi ''(t)+\left[-2\xi t^3+2\nu t^2+2\xi\beta^2t+\beta^2\right]\chi '(t)-\left[(2\xi\nu-\lambda_1)t^2+\lambda_2t+\xi\beta^2\right]\chi(t)=0.
\end{equation}
These equations are now suitable for the application of the BAM. 
\section{The Bethe Ansatz solutions to the basic equation}
It can be seen that Eqs.\,\eqref{eq:16}, \eqref{eq:17}, \eqref{eq:26} and \eqref{eq:27} have a basic form 
\begin{equation}\label{eq:28}
\left[P(t)\frac{d^2}{dt^2}+Q(t)\frac{d}{dt}+R(t)\right]S(t)=0,
\end{equation}
where
$P(t), Q(t)$ are polynomials of at most degree 3 and $R(t)$ is a polynomial of at most degree 2, which we write as
\begin{equation}\label{eq:29}
P(t)=\sum_{k=0}^3a_kt^k,\hspace{0.2in}Q(t)=\sum_{k=0}^3b_kt^k,\hspace{0.2in}R(t)=\sum_{k=0}^2c_kt^k.
\end{equation}
where $a_k, b_k$ and $c_k$ are constants. This basic equation is quasi-exactly solvable for certain values of its parameters and exact solutions are given
by polynomials of degree $n$ in $t$ with $n$ being non-negative integers. 
In fact, the above equation is a special case of the general 2nd order differential equations solved in \cite{27} by means of the Bethe ansatz method. Applying the results in \cite{27}, we have \\\\
{\bf Proposition 3.1} {\it Given a pair of polynomials $P(t)$ and $Q(t)$, then the values of the coefficient $c_0, c_1, c_2$ of polynomial $R(t)$ such that the differential equation \eqref{eq:28} has a degree $n$ polynomial solution 
\begin{equation}\label{eq:30}
S(t)=\prod_{i=1}^n(t-t_i),\hspace{0.1in}S(t)\equiv 1\hspace{0.1in}\mbox{for}\hspace{0.1in} n=0
\end{equation}
 with distinct roots $t_1, t_2,\dots,t_n$ are given by 
\begin{equation}\label{eq:31}
-c_2=nb_3,
\end{equation}
\begin{equation}\label{eq:32}
-c_1=b_3 \sum_{i=1}^nt_i+n(n-1)a_3+nb_2,
\end{equation}
\begin{equation}\label{eq:33}
-c_0=b_3 \sum_{i=1}^nt_i^2+\left[2(n-1)a_3+b_2\right]\sum_{i=1}^nt_i+n(n-1)a_2+nb_1,
\end{equation}
where the roots $t_1,t_2,\dots,t_n$ satisfy the Bethe ansatz equations
\begin{equation}\label{eq:34}
\sum_{j\neq i}^n\frac{2}{t_i-t_j}+\frac{b_3t_i^3+b_2t_i^2+b_1t_i+b_0}{a_3t_i^3+a_2t_i^2+a_1t_i+a_0}=0,\hspace{0.1in}i=1,2,\dots,n
\end{equation}
The above equations \eqref{eq:31}--\eqref{eq:34} give all polynomial $R(t)$ such that the ODE \eqref{eq:28} has a polynomial solution \eqref{eq:30}.}

One major task in the application of BAM is obtaining the roots of the $n$ algebraic Bethe ansatz equations \eqref{eq:34}. For an arbitrary $n$, the equation is very difficult, if not impossible, to solve algebraically. However, numerical solutions to the Bethe ansatz equations have also been discussed in many applications \cite{31}-\cite{35}. In what follows, we give exact algebraic solutions for each of the model using the above results.

\section{Solutions to Relativistic Soft-core Coulomb Models}
In this section, we apply the results in section 3 above to obtain the energies and the wavefunctions of the models in section 2. By using appropriate equations, we also work out the possible constraints on the potential parameters.

\subsection{Solutions to Dirac equation with the Soft-core Coulomb potentials}
\subsubsection{Case $q=1$}
By comparing Eqs.\,\eqref{eq:16} and \eqref{eq:28}, we have $a_3=a_0=b_3=b_0=0$, $a_2=1$, $a_1=-\beta$, $b_2=-2\sigma$, $b_1=2(\kappa+\sigma\beta)$, $c_1=-(2\sigma\kappa+\gamma Z_\delta)$ and $c_0=\gamma Z_\delta\beta$. Thus, by Eqs.\,\eqref{eq:32} and \eqref{eq:33}, we have the energy equation
\begin{equation}\label{eq:39}
2\sigma(n+\kappa)=\gamma Z_\delta
\end{equation}
and
\begin{equation}\label{eq:40}
n(n+2\kappa+2\sigma\beta-1)-2\sigma\sum_{i=1}^nt_i=-\gamma Z_\delta\beta
\end{equation}
respectively, provided the the roots $\{t_i\}$ satisfy the Bethe ansatz equation 
\begin{equation}\label{eq:41}
\sum_{i\neq j}^n\frac{1}{t_i-t_j}+\frac{-\sigma t_i^2+(\kappa+\sigma\beta)t_i}{t_i(t_i-\beta)}=0. 
\end{equation}
Using Eqs.\,\eqref{eq:12} and \eqref{eq:39}, one can obtain the relativistic energy (Dirac case) for the family of soft core potentials 
\begin{equation}\label{eq:42}
E_{q,n}(\kappa)=\frac{(\mu+C_q)Z_\delta^2-4\mu(n+\kappa)^2}{4(n+\kappa)^2+Z_\delta^2}.
\end{equation}
For the first excited state(corresponding to  $n=1$), Eq.\,\eqref{eq:41} becomes
\begin{equation}\label{eq:43}
t_1(-\sigma t_1+\kappa+\sigma\beta)=0\hspace{0.1in}\Rightarrow\hspace{0.1in}t_1=0\hspace{0.1in}\mbox{and}\hspace{0.1in}t_1=\frac{\kappa+\sigma\beta}{\sigma}.
\end{equation}
Substitution these solutions into Eq.\,\eqref{eq:40} we have the condition when $t_1=0$
\begin{equation}\label{eq:44}
\beta(2\sigma+\gamma Z_\delta)+2\kappa=0.
\end{equation}
The other solution yields $\beta=0$ and hence discarded. The spinor wavefunctions of the first excited state can then be written as
\begin{equation}\label{eq:45}
\left(\begin{array}{lll}
F_{1,1}(r)\\\\
G_{1,1}(r)
\end{array}\right)\sim r^\kappa e^{-\sigma(r+\beta)}\left(\begin{array}{lll}
\frac{1-\sigma(r+\beta)}{\mu-E_{1,1}+C_1}\\\\
\hfil r+\beta
\end{array}\right)
\end{equation}
with
$$\sigma=\frac{\gamma Z_\delta}{2(\kappa+1)}.$$
where we have used Eq.\,\eqref{eq:7b} to obtain $F_{1,1}(r)$. 

Also, the second excited state(corresponding to $n=2$), Eq.\,\eqref{eq:40} gives the following set of equations
\begin{equation}\label{eq:46}
\frac{t_1(t_1-\beta)}{t_1-t_2}+t_1(-\sigma t_1+\kappa+\sigma\beta)=0$$
$$\frac{t_2(t_2-\beta)}{t_2-t_1}+t_2(-\sigma t_2+\kappa+\sigma\beta)=0,
\end{equation}
with non-identical sums of roots
$$t_1+t_2=\frac{2\kappa+2\sigma\beta+1}{\sigma}$$
\begin{equation}\label{eq:47}
t_1+t_2=\frac{(1+\kappa+\sigma\beta)\pm\sqrt{(1+\kappa+\sigma\beta)^2-4\sigma\beta}}{2\sigma}.
\end{equation}
Substituting these solutions into Eq.\,\eqref{eq:40}, the first solution gives $\beta=0$ and hence discarded, while the other gives the following condition 
\begin{equation}\label{eq:48}
\gamma^2Z_\delta^2\beta^2(\kappa^2+7\kappa+11)+2\gamma Z_\delta\beta(3\kappa^3+15\kappa^2+22\kappa+8)+4\kappa(\kappa+1)(\kappa+2)^2=0
\end{equation}The spinors components of the second excited state can then be written as 
\begin{equation}\label{eq:49}
\left(\begin{array}{lll}
F_{1,2}(r)\\\\
G_{1,2}(r)
\end{array}\right)\sim r^\kappa e^{-\sigma(r+\beta)}\left(\begin{array}{lll}
\frac{(r+\beta-\varrho)+(r+\beta)(1-\sigma(r+\beta-\varrho))}{\mu-E_{1,2}+C_1}\\\\
\hfil (r+\beta)(r+\beta-\varrho)
\end{array}\right)
\end{equation}with
\begin{equation}\label{eq:50}
\varrho=\frac{(1+\kappa+\sigma\beta)\pm\sqrt{(1+\kappa+\sigma\beta)^2-4\sigma\beta}}{2\sigma}
\end{equation}
and
\begin{equation}\label{eq:51}
\sigma=\frac{\gamma Z_\delta}{2(\kappa+2)}.
\end{equation}
\subsubsection{Case $q=2$}
By comparing Eqs.\,\eqref{eq:17} and \eqref{eq:28}, we have $a_2=a_0=0$, $a_3=1$, $a_1=-\beta^2$, $b_3=-2\sigma$, $b_2=2\kappa$, $b_1=2\sigma\beta^2$, $b_0=\beta^2$, $c_2=-(2\sigma\kappa+\gamma Z_\delta)$, $c_1=-\sigma^2\beta^2$ and $c_0=-\sigma\beta^2$. %It is clear by Eqs.\,\eqref{eq:31} that both potentials $V_1$ and $V_2$ share the same energy spectrum. 
Then by \eqref{eq:32} and \eqref{eq:33}, we have the following equations
\begin{equation}\label{eq:53}
n(n+2\kappa-1)-2\sigma\sum_{\alpha=1}^nt_\alpha=\sigma^2\beta^2,
\end{equation}
\begin{equation}\label{eq:54}
(n+\kappa-1)\sum_{i=1}^nt_i-\sigma\sum_{i=1}^nt_i^2=-\sigma\beta^2\left(n-\frac{1}{2}\right),
\end{equation}
respectively, with $\{t_i\}$ satisfying the Bethe ansatz equation
\begin{equation}\label{eq:55}
\sum_{i\neq j}^n\frac{2}{t_i-t_j}+\frac{-2\sigma t_i^3+2\kappa t_i^2+2\sigma\beta^2t_i+\beta^2}{t_i(t_i^2-\beta^2)}=0. 
\end{equation}
%It is interesting to note that for both cases, i.e $q=1$ and $q=2$, the energy equals that of the Coulomb potential. 
For the first excited state (corresponding to $n=1$), Eq.\,\eqref{eq:55} becomes 
\begin{equation}\label{eq:56}
-2\sigma t_1^3+2\kappa t^2_1+2\sigma \beta^2t_1+\beta^2=0,
\end{equation}
with real solution
\begin{equation}\label{eq:57}
t_1=\frac{1}{3\sigma }\left(\frac{A}{2}+\frac{2(3\beta^2\sigma ^2+\kappa^2)}{A}+\kappa\right),
\end{equation}
where
\begin{equation}\label{eq:58}
A=\left(36\beta^2\sigma ^2\kappa+54\beta^2\sigma ^2+8\kappa^3+6\beta\sqrt{-48\beta^4\sigma ^4-12\beta^2\sigma ^2\kappa^2+108\beta^2\sigma ^2\kappa+81\beta^2\sigma ^2+24\kappa^3\sigma }\right)^{\frac{1}{3}}
\end{equation}
Substituting $t_1$ into Eq.\,\eqref{eq:54}, and solving the resulting equation, we have the non-zero solution 
\begin{equation}\label{eq:59}
\sigma \beta=\pm\sqrt{ 2(\kappa+1)}\hspace{0.1in}\Rightarrow\hspace{0.1in}\gamma^2Z_\delta^2\beta^2=(2\kappa+2)^3.
\end{equation}
Thus, the spinors wavefunction for the first excited state is therefore given as 
\begin{equation}\label{eq:61}
\left(\begin{array}{lll}
F_{2,1}(r)\\\\
G_{2,1}(r)
\end{array}\right)\sim r^\kappa e^{-\sigma\sqrt{r^2+\beta^2}}\left(\begin{array}{lll}
\frac{r}{\Lambda\sqrt{r^2+\beta^2}}-\frac{\sigma r\left(\sqrt{r^2+\beta^2}-(2\kappa-\sigma^2\beta^2)/2\sigma^2\right)}{\Lambda\sqrt{r^2+\beta^2}}\\\\
\hfil \sqrt{r^2+\beta^2}-(2\kappa-\sigma^2\beta^2)/2\sigma^2
\end{array}\right)
\end{equation}
with
$\Lambda=\mu-E_{2,1}+C_2$ and
$$\sigma=\frac{\gamma Z_\delta}{2(\kappa+1)}.$$
\subsection{Solutions to Klein-Gordon equation with the Soft-core Coulomb \newline potentials}
\subsubsection{Case $q=1$}
By comparing Eqs.\,\eqref{eq:26} and \eqref{eq:28}, we have $a_1=a_0=b_1=b_0=0$, $a_3=1$, $a_2=-\beta$, $b_3=-2\xi$, $b_2=2(\nu+\xi\beta)$, $c_2=-(2\xi\nu-\lambda_1)$, $c_1=-(\lambda_1\beta+\lambda_2)$ and $c_0=\lambda_2\beta$. Thus, by Eqs.\,\eqref{eq:31}, \eqref{eq:32} and \eqref{eq:33}, we have the following equations
\begin{equation}\label{eq:62}
2\xi(\nu+n)=\lambda_1,
\end{equation}
\begin{equation}\label{eq:63}
n(n+2\nu+2\xi\beta-1)-2\xi\sum_{i=1}^nt_i=\lambda_1\beta+\lambda_2,
\end{equation}
\begin{equation}\label{eq:64}
(n+\nu+\xi\beta-1)\sum_{i=1}^nt_i-\xi\sum_{\alpha=1}^nt_i^2=\frac{\beta[n(n-1)-\lambda_2]}{2}
\end{equation}
provided the roots $t_i$ satisfy the Bethe ansatz equations
\begin{equation}\label{eq:65}
\sum_{i\ne j}^n\frac{1}{t_i-t_j}+\frac{t^2_i\left(-\xi t_i+\nu+\xi\beta\right)}{t^2_i(t_i-\beta)}=0.
\end{equation}
With the help of Eqs.\,\eqref{eq:22} and \eqref{eq:62}, we obtain the relativistic energy (Klein-Gordon Case)
\begin{equation}\label{eq:66}
E_{n}=-\mu\frac{Z_vZ_s+(n+\nu)\sqrt{Z_v^2-Z_s^2+(n+\nu)^2}}{Z_v^2+(n+\nu)^2}
\end{equation}
The above energy equation is independent of the parameter $q$, thus the family of the soft-core potentials has a common spectrum for the Klein-Gordon case.

For the first excited state(corresponding to $n=1$), Eqs.\,\eqref{eq:63}, \eqref{eq:64} and \eqref{eq:65} yield
$$2(\nu+\xi\beta)-2\xi t_1=\lambda_1\beta+\lambda_2,$$
$$(\nu+\xi\beta)t_1-\xi t_1^2=-\beta\lambda_2/2$$ and
$$t_1=0\hspace{0.1in}\mbox{and}\hspace{0.1in}t_1=\frac{\nu+\xi\beta}{\xi},$$
with possible solutions $\{t_1=0,\lambda_2=0,\beta=2(\nu+1)/\lambda_1\}$ and $\{t_1=(\nu+\xi\beta)/\xi,\lambda_2=0,\newline\beta=0\}$, where we have used the fact that for $n=1$, $\xi=\lambda_1/2(\nu+1)$. Thus, with the choice of the first set, we write the wavefunction of the first excited state as follows
\begin{equation}\label{eq:67}
\phi_{1,1}(r)\sim r^{\ell+1}e^{-\xi(r+\beta)}(r+\beta).
\end{equation}
Similarly, for second excited state ($n=2$), we have the following non-identical sums of roots from the Bethe ansatz equation \eqref{eq:65}
\begin{equation}\label{eq:68}
t_1+t_2=\frac{2\nu+2\xi\beta+1}{\xi}$$
$$t_1+t_2=\frac{(1+\nu+\xi\beta)\pm\sqrt{(1+\nu+\xi\beta)^2-4\xi\beta}}{2\xi}.
\end{equation}
On substituting into Eqs.\,\eqref{eq:63} and \eqref{eq:64}, the first sum gives $\beta=\lambda_2=0$ and hence discarded, while the second sum gives $\lambda_2=0$ and the following condition 
\begin{equation}\label{eq:69}
\nu(\nu+1)\beta^2\lambda_1^2-(\nu+2)(6\nu^2+4\nu-1)\beta\lambda_1+(\nu+2)^2(8\nu^2-2\nu-1)=0.
\end{equation}
Thus the wavefunction of the second excited state can be written as
\begin{equation}\label{eq:70}
\phi_{1,2}(r)\sim r^{\ell+1}(r+\beta)e^{-\xi(r+\beta)}(r+\beta-\rho),
\end{equation} where
\begin{equation}\label{eq:71}
\rho=\frac{(2+\ell+\xi\beta)\pm\sqrt{(2+\ell+\xi\beta)^2-4\xi\beta}}{2\xi}
\end{equation}
In both cases above, the condition on parameter $\lambda_2$ reveals that the system has an exact polynomial solution only for the case of equal vector and scalar potentials.
\subsubsection{Case $q=2$}
By comparing Eqs.\,\eqref{eq:27} and \eqref{eq:28}, we have $a_2=a_0=0$, $a_3=1$, $a_2=-\beta^2$, $b_3=-2\xi$, $b_2=2\nu$, $b_1=2\xi\beta^2$, $b_0=\beta^2$, $c_2=-(2\xi\nu-\lambda_1)$, $c_1=-\lambda_2$ and $c_0=-\xi\beta^2$. Thus, by Eqs.\,\eqref{eq:31}, \eqref{eq:32} and \eqref{eq:33}, we have the following equations
\begin{equation}\label{eq:72}
2\xi(\nu+n)=\lambda_1,
\end{equation}
\begin{equation}\label{eq:73}
n(n+2\nu-1)-2\xi\sum_{i=1}^nt_i=\lambda_2,
\end{equation}
\begin{equation}\label{eq:74}
(n+\nu-1)\sum_{i=1}^nt_i-\xi\sum_{i=1}^nt_i^2=-\xi\beta^2\left(n-\frac{1}{2}\right)
\end{equation}
provided the roots $t_i$ satisfy the Bethe ansatz equations
\begin{equation}\label{eq:75}
\sum_{i\ne j}^n\frac{1}{t_i-t_j}+\frac{-2\xi t^3_i+2\nu t_i^2+2\xi\beta^2t_i+\beta^2}{t_i(t^2_i-\beta^2)}=0.
\end{equation}
%From Eq.\,\eqref{eq:72},both potentials have the same energy spectum as expected. 
For the first excited state ( $n=1$), the Bethe ansatz equation \eqref{eq:75} becomes
\begin{equation}\label{eq:76}
-2\xi t_1^3+2\nu t^2_1+2\xi\beta^2t_1+\beta^2=0,
\end{equation}
with real-valued solution
\begin{equation}\label{eq:77}
t_1=\frac{1}{3\xi}\left(\frac{A}{2}+\frac{2(3\beta^2\xi^2+\nu^2)}{A}+\nu\right),
\end{equation}
where
\begin{equation}\label{eq:78}
A=\left(36\beta^2\xi^2\nu+54\beta^2\xi^2+8\nu^3+6\beta\xi\sqrt{-48\beta^4\xi^4-12\beta^2\xi^2\nu^2+108\beta^2\xi^2\nu+81\beta^2\xi^2+24\nu^3}\right)^{\frac{1}{3}}
\end{equation}
Substituting $t_1$ into Eq.\,\eqref{eq:73} and \eqref{eq:74} and using \eqref{eq:72}, we have the following conditions 
\begin{equation}\label{eq:79}
\lambda_2=2\nu-2\xi t_1
\end{equation}
and
\begin{equation}\label{eq:80}
\xi\beta=\pm\sqrt{ 2(\nu+1)}\hspace{0.1in}\Rightarrow\hspace{0.1in}\beta^2\lambda_1^2=8(\nu+1)^3.
\end{equation}
The wavefunction for the first excited state is therefore given as 
\begin{equation}\label{eq:81}
\Psi_{2,1}(r)\sim r^{\ell+1}e^{-\xi\sqrt{r^2+\beta^2}}\left(\sqrt{r^2+\beta^2}-t_1\right),
\end{equation}
where
$$\xi=\frac{\lambda_1}{2(\nu+1)}$$
and $\nu=\ell+1$.
\section{Hidden Algebraic Structure}
One way to understand the QES theory is to demonstrate that the Hamiltonian  can be expressed in terms of generator of a Lie algebra
\begin{equation}\label{eq:82}
J^-=\frac{d}{dt},\hspace{0.2in}J^+=t^2\frac{d}{dt}-nt,\hspace{0.2in}J^0=t\frac{d}{dt}-\frac{n}{2},
\end{equation}
which are differential operator realization of the $n+1$ dimensional representation of the $sl(2)$ algebra. Moreover, if  we write the basic equation \eqref{eq:28} in the Schr\"odinger form
\begin{equation}\label{eq:83}
HS(t)=-c_0S(t)
\end{equation}
where $-c_0$ is the eigenvalue of the Hamiltonian $H$, then it can easily be shown that if $c_1=-nb_2$, with $n$ being any non-negative integer, the differential operator $H$ is an element of the enveloping algebra of Lie algebra $sl(2)$
\begin{equation}\label{eq:84}
H=a_3J^+J^0+a_2J^0J^0+a_1J^0J^-+a_0J^-J^-+\left[\frac{1}{2}(3n-2)a_3+b_2\right]J^+$$$$+\left[(n-1)a_2+b_1\right]J^0+\left(\frac{n}{2}a_1+b_0\right)J^--\frac{n^2}{4}a_2+\frac{n}{2}\left[(n-1)a_2+b_1\right]
\end{equation}
provided 
\begin{equation}\label{eq:85}
b_3=c_2=0\hspace{0.1in}\mbox{and}\hspace{0.1in}c_1=-n\left[(n-1)a_3+b_2\right].
\end{equation} Except for the case $q=1$ of the Dirac equation, (Eq.\,\eqref{eq:16}), all other case violate the condition \eqref{eq:85}. Thus for Eq.\,\eqref{eq:16} we have 
\begin{equation}\label{eq:86}
{H}=t(t-\beta)\frac{d^2}{dt^2}+2[(\kappa+\sigma\beta)t-\sigma t^2)]\frac{d}{dt}-[(2\sigma\kappa+\gamma Z_\delta)t]
\end{equation} and 
\begin{equation}\label{eq:87}
c_0=\gamma Z_\delta\beta
\end{equation} with $sl(2)$ algebraization
\begin{equation}\label{eq:88}
\small H=J^0J^0-\beta J^0J^--2\sigma J^++\left[(n-1)+2(\kappa+\sigma\beta)\right]J^0-\frac{n\beta}{2}J^-+\frac{n}{4}\left[n+2(\kappa+\sigma\beta-1)\right],
\end{equation}
provided the energy equation \eqref{eq:39} is satisfied.

\section{Concluding Remarks}
By considering a unified case, we have constructed the Bethe ansatz solutions to relativistic soft-core Coulomb models, within the frame work of the Dirac and Klein-Gordon equations. We showed that in both cases, the equations are reducible to the same basic equation which has an exact solution, provided the parameters satisfy certain constraints. Unlike previous non-relativistic cases \cite{9,10,11}, the quasi-exact solvability of the equations has enabled us to use proposition 3.1 to obtain closed form expressions for the energies and eigenfunctions.

Moreover, we have also shown that except for the Dirac equation with the shifted Coulomb model ($q=1$), there are no $sl(2)$ algebraic structures in the relativistic soft-core Coulomb models. Thus these models provide interesting examples of QES systems without $sl(2)$ algebraization. Let us remark at this point that the existence of a underlying Lie algebraic 
structure in a differential equation is only a sufficient condition for the differential equation to be 
quasi-exactly solvable. In fact there are more general (than the Lie-algebraically based)
differential equations which do not possess a underlying Lie algebraic structure but are
nevertheless quasi-exactly solvable (i.e have exact polynomial solutions ) \cite{27,36}.

 Finally, it is pertinent  to note that our method gives a more general closed form expressions for the solutions, however the determination of the roots of the Bethe ansatz equations for higher excited states may be a major difficulty in the application of the method.   
\section*{Acknowledgements}
This work was supported by the Australian
IPRS and a University of Queensland Centennial Scholarship. YZZ is supported in part by the Australian
Research Council through Discovery Project DP110103434. DA is indebted to Father J and Agboola B for
their support during the preparation of the manuscript. We also wish to thank the referees for their very useful comments.

\end{document}